\documentclass[aps,prc,twocolumn,superscriptaddress,showpacs,floatfix]{revtex4-1}
\usepackage{graphicx,amssymb}
\usepackage[fleqn]{amsmath}
\usepackage{amsfonts}
\usepackage{dcolumn}% Align table columns on decimal point
\usepackage{bm}% bold math
\usepackage{braket}
\usepackage{multirow} 
\usepackage{MnSymbol}
\usepackage{color}
\usepackage{hyperref}

\begin{document}
%\begin{CJK*}{UTF8}{gbsn}

\date{\today}

%\title{Energy Spectrum of Neutron-Rich Helium Isotopes: Complex Made Simple} % 10 words
\title{Energy Spectrum of Neutron-Rich Helium Isotopes: Complex Made Simple} % 10 words

\author{K. Fossez}
\affiliation{NSCL/FRIB Laboratory,
Michigan State University, East Lansing, Michigan 48824, USA}

\author{J. Rotureau}
\affiliation{NSCL/FRIB Laboratory,
Michigan State University, East Lansing, Michigan 48824, USA}

\author{W. Nazarewicz}
\affiliation{Department of Physics and Astronomy and NSCL/FRIB Laboratory,
Michigan State University, East Lansing, Michigan 48824, USA}

\begin{abstract}

	We demonstrate that the intricate energy spectrum of neutron-rich helium isotopes can be straightforwardly described by taking advantage of the low-energy properties of neutron-neutron interaction and the scale separation that is present in diluted dripline systems. By using arguments based on the halo effective field theory, we carry out a parameter reduction of the complex-energy configuration interaction framework in the $spd$ space, including resonant and scattering states. By constraining the core potential to $\alpha$-n scattering phase-shifts and adjusting the strength of the spin-singlet central neutron-neutron interaction, we reproduce experimental energies and widths of $^{5-8}$He within tens of keV precision. We predict a parity inversion of narrow resonances in $^{9}$He and show that the ground state of $^{10}$He is an $s$-wave-dominated configuration that could decay through two-neutron emission. This threshold state can be viewed as a ``double-halo" structure in an analogy to the atomic ${ {}^{3}\text{He}{}^{4}\text{He}_{2} }$ trimer.
\end{abstract}

\maketitle

%%%%%%%%%%%%%%%%%%%%%%%%%%%%%%%%%%% intro %%%%%%%%%%%%%%%%%%%%%%%%%%%%%%%%%%%

\begin{figure}[htb]
	\includegraphics[width=1.0\linewidth]{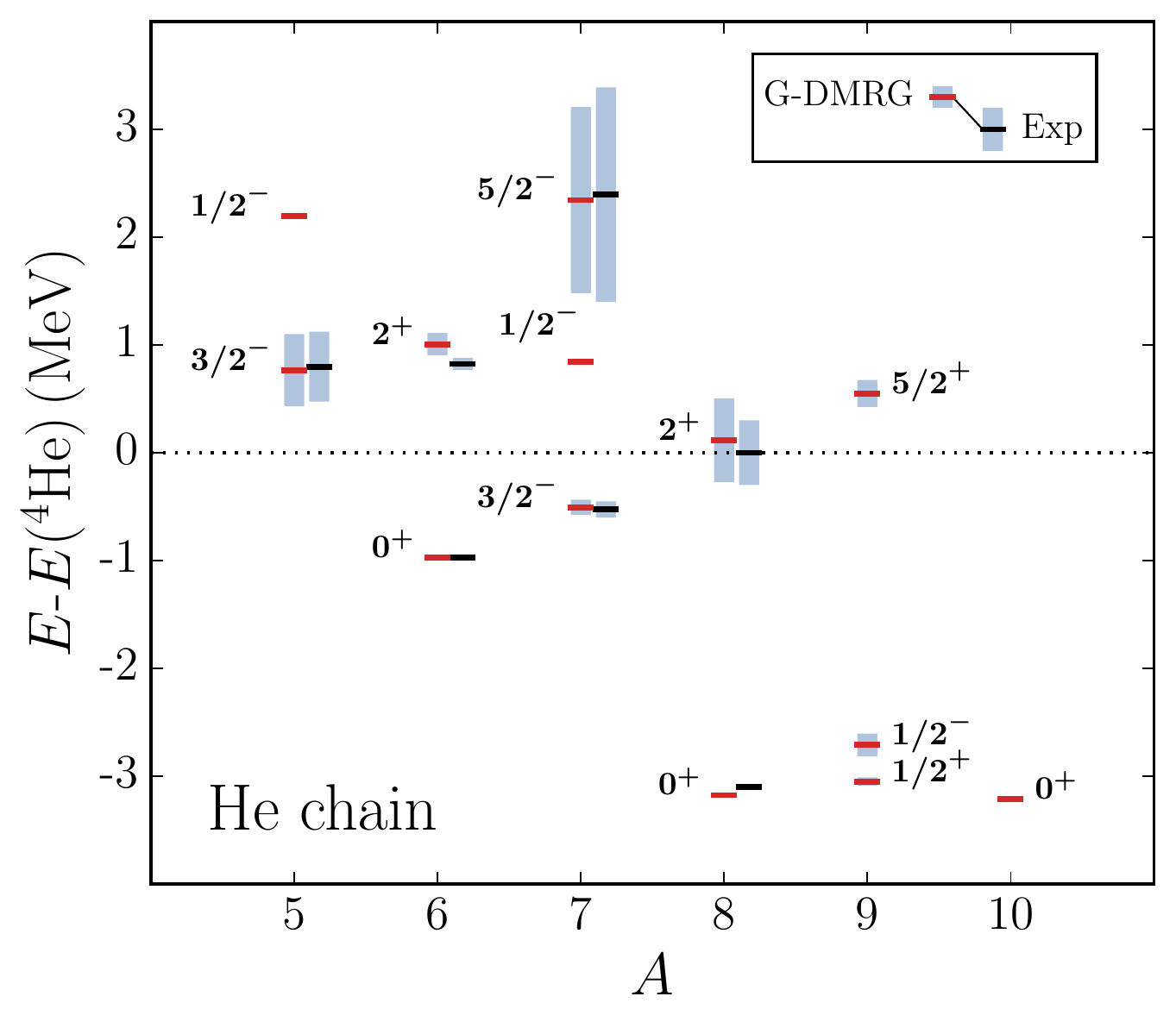}%plot_He_chain_levels_5.pdf
	\caption{Energy spectra of $^{5-10}$He with respect to the $^{4}$He core. Experimental data \cite{ensdf} are compared to our Gamow-DMRG (G-DMRG) calculations. Decay widths are shown as shaded bars. The predicted $1/2^-$ resonant states in $^{5,7}$He are so broad that their widths are not marked. For these states, as well as for states in $^{9,10}$He, experimental information is not firm.}
	\label{spectrum}
\end{figure}

\textit{Introduction}--The neutron-rich helium isotopes $^{5-10}$He epitomize novel aspects of nuclear structure at and beyond the limit of nuclear binding. Experimentally, the even-even isotopes $^6$He \cite{tilley02_1931} and $^{8}$He \cite{tanihata85_1604,golovkov09_1537} are Borromean halos, they have no bound excited states, and they exhibit an abnormal pattern of the one- and two-neutron emission thresholds. The odd-$N$ isotopes $^{5}$He \cite{kobayashi97_1913,aleksandrov98_1918,tilley02_1931} and $^{7}$He \cite{denby08_1536,aksyutina09_1888} are neutron-unbound. Presently, too little is known about the elusive $^{9}$He \cite{Tilley2004,kalanee13_1909} and $^{10}$He \cite{korsheninnikov94_1914,ostrowski94_1915,kobayashi97_1913,golovkov09_1537,johansson10_1890,sidorchuk12_1912,sharov14_1911,aoyama02_1923,grigorenko08_1921,jones15,fortune13_1910,fortune15_1869} isotopes to firmly conclude whether they represent genuine nuclear systems or not. The current experimental information on the energy spectrum of $^{5-10}$He is displayed in Fig.~\ref{spectrum}.
%%%%
%%%%

Theoretically, the understanding of the neutron-rich helium chain is challenging: it requires a microscopic framework based on a realistic interaction that is capable of describing many-body correlations and salient continuum effects \cite{forssen13_394,nazarewicz16_1810}. A number of sophisticated many-body methods, based on realistic Hamiltonians, were employed to describe neutron-rich helium isotopes using nucleons as elementary degrees of freedom, both without continuum couplings \cite{pudliner97_924,caurier06_1924,lisetskiy08_933,saaf14} and also considering them to some extent \cite{nollett07_944,hagen07_976,nollett12_835,papadimitriou13_441,baroni13_753,baroni13_385,vorabbi18_1977}. While such $A$-body approaches are powerful, they also have shortcomings when it comes to quantitative and quantified predictions. The associated two- and three-body forces, often derived from chiral effective field theory \cite{kolck94_1147,epelbaum09_866,machleidt11_414}, are in most cases not statistically optimized and quantified; hence they do not have the required precision and accuracy to guide experiments on exotic nuclei near the drip lines \cite{fossez17_1927}. It also remains to be seen how the truncation errors at the two- and three-body level \cite{furnstahl15_1746,furnstahl15_1354,melendez17_1978} would propagate in many-body calculations. Moreover, the complete inclusion of many-body forces is still computationally challenging, which only adds to the already difficult task of including continuum couplings. Consequently, no satisfactory $A$-body description of $^{9,10}$He has thus been achieved.

In this paper, we demonstrate that it is possible to achieve a precise description of the neutron-rich helium chain within an effective framework that recognizes the emergence of effective scales and associated degrees of freedom in these nuclei. We first note that, at low-energy, the tightly bound nature of $^{4}$He makes it a natural core whose internal dynamics is largely decoupled from valence neutrons. This decoupling is reflected in the smallness of the ratio $|S_{\rm 1n}|$($^5$He)/$E^*$($^4$He, $0^+_2$)$\approx$0.04. This makes it possible to reduce the full $A$-body neutron-rich helium problem to a reduced-size task involving the $^4$He core and $(A-4)$ neutrons. As $^{6,8}$He are halo systems, further simplifications are possible by taking advantage of the scale separation. For instance, halo effective field theory (halo-EFT) \cite{bedaque02_1957,hammer17_1959} allows to systematically construct, order by order, effective interactions tailored to weakly bound systems \cite{bertulani02_869,bedaque03_1085,rotureau13_207,ji14_1101}. 

According to the power counting in pionless EFT \cite{bedaque02_1957,kolck99_1979,chen99_1053}, the dominant contribution to the neutron-neutron interaction at low energy should come from the $(S=0, L=0)$ channel, while the contributions from channels with $L \geq 1$ should only appear at higher orders. Unfortunately, many-body terms appearing in the halo-EFT Hamiltonian, which are already present at the lowest order in $^6$He in the form of a $^4$He-neutron-neutron interaction \cite{rotureau13_207,ji14_1101}, make this approach unpractical when it comes to the heaviest neutron-rich helium isotopes. Still, the recognition that the main contribution to the valence neutron-neutron interaction in the neutron-rich helium isotopes primarily comes from the $^1S_0$ channel, suggests that rather simple interactions should perform well in those diluted many-neutron systems. This has been recognized in the studies of neutron drops~\cite{Gandolfi2017}. Indeed, because of the dilute character of those systems, the role of many-body interactions is expected to be small. As shown in Ref.~\cite{Gandolfi2017}, the ground-state (g.s.) energy pattern of trapped neutron drops is strikingly reminiscent of that for the helium chain.

The strategy based on the $^{4}$He core was adopted by continuum shell model approaches to describe the helium chain \cite{michel04_93,volya05_470,rotureau06_15,papadimitriou11_277,jaganathen17_1974}. All these approaches relied on phenomenological one- and two-body interactions in the valence neutron space, which were not constructed using effective scale arguments, and, except for the recent work in Ref.~\cite{jaganathen17_1974}, no systematic study of the model parameter space has been carried out. Moreover, in all previous shell-model studies, the continuum space pertaining to the unbound $^{9,10}$He isotopes has been truncated. The fact that none of the traditional approaches, whether $A$-body methods or shell model approaches, are either practical or can provide reliable predictions for neutron-rich helium isotopes motivates the development of an alternative path rooted in halo-EFT and based on the complex-energy formalism. 

%%%%%%%%%%%%%%%%%%%%%%%%%%%%%%%%%%% method %%%%%%%%%%%%%%%%%%%%%%%%%%%%%%%%%%%

\textit{Method}--In the present work, the description of neutron-rich helium isotopes is achieved by employing the single-particle (s.p.) Berggren basis \cite{berggren68_32,berggren93_481}. The use of the Berggren ensemble allows to naturally extend the configuration-interaction picture into the complex-energy plane \cite{michel09_2}, by explicitly including Gamow (resonant) states and nonresonant scattering states for each partial-wave channel ${ c = ( \ell , j ) }$. As discussed in detail in Ref.~\cite{michel09_2} in the context of complex-energy shell model applications, scattering states entering the Berggren basis are defined along a contour ${ \mathcal{L}_{c}^{+} }$ in the fourth quadrant of the complex-momentum plane that surrounds the resonant poles ${ \{ {k}_{i} \} }$ and then extends to ${ k \to +\infty }$. In practice, the integral along the contour ${ \mathcal{L}_{c}^{+} }$ is discretized using a Gauss-Legendre quadrature, and then a many-body basis made of Slater determinants can be constructed as usual.

The numerical resolution of the many-body problem is performed using the density matrix renormalization group (DMRG) method for open quantum systems \cite{rotureau06_15,rotureau09_140} or Gamow-DMRG (G-DMRG), which has been shown to be a powerful technique to handle large many-body spaces. Also, working within a basis generated with natural orbitals \cite{brillouin33} allows to significantly speed-up the numerical convergence of the G-DMRG method \cite{shin16_1860,fossez16_1793,fossez17_1916}. 

Our strategy is to make a parameter reduction of the G-DMRG Hamiltonian using effective scale arguments. The goal is to rearrange Hamiltonian terms similarly to what is usually done in core-based shell model approaches or, more microscopically, using the in-medium similarity renormalization group approach \cite{tsukiyama12_1151,hergert16_1673}. 
% effective SM interaction from coupled cluster
% jansen14_1150
First, the one-body $^{4}$He-neutron interaction is taken in a Woods-Saxon (WS) form. It contains the central and spin-orbit terms, whose parameters were optimized to the $s$ and $p$ phase shifts in the $\alpha-n$ scattering \cite{hoop66_989,stammbach72_1942,bond77_1941} as was done in Refs.~\cite{varga94_2024,theeten06_2025,jaganathen17_1974}. The resulting WS parameters are: the depth ${ {V}_{0} = 41.77 }$ MeV, the diffuseness ${ a = 0.618 }$ fm, the radius ${ {R}_{0} = 2.162 }$ fm, and the spin-orbit strength ${ {V}_{\text{so}} = 6.991 }$ MeV. By construction, the energies and widths of the ${ {J}^{\pi} = {3/2}^{-} }$ ground state and the ${ {J}^{\pi} = {1/2}^{-} }$ broad excited state of $^{5}$He are reproduced, with the latter being solely a pole of the $S$-matrix rather than a genuine resonance. These parameters coincide within the error bars with those obtained in the recent optimization study \cite{jaganathen17_1974}. This choice of the one-body potential departs from halo-EFT but provides a simple way to include $\alpha-n$ correlations.

In a second step, we reduce the interaction between valence neutrons to a residual two-body force using insights from halo-EFT. 
For the two-neutron interaction, we take a reduced variant of the Furutani-Horiuchi-Tamagaki (FHT) interaction \cite{furutani78_1012,furutani79_1013}. In this interaction, four terms are present in the isovector channel: two central terms in the spin-singlet and spin-triplet channels, and one spin-orbit term and one tensor term in the spin-triplet channel. However, based on the halo-EFT argument, we reduce the FHT interaction to the single central term in the spin-singlet channel. We note in passing that this argument explains the sloppiness of the parameters associated with the spin-triplet channels seen in Ref.~\cite{jaganathen17_1974}. The leading-order of halo-EFT \cite{bertulani02_869,bedaque03_1085} involves the $^1S_0$ channel only. Here we also consider the $L>0$ spin-singlet channels to be able to check \textit{a posteriori} that the main contribution comes from the $^1S_0$ channel.

The form factor for the central FHT term is a sum of three Gaussians with different ranges: ${ ({r}_{0} = 0.160 \, \text{fm} )}$, ${ ({r}_{1} = 1.127 \, \text{fm} )}$ and ${ ({r}_{2} = 3.400 \, \text{fm} )}$. This is another difference with halo-EFT at leading-order as in the latter case the interaction is given by a regularized delta force in the $^1S_0$ channel \cite{bertulani02_869,bedaque03_1085}, which can be taken in a single Gaussian form. We stick to the original FHT form factor as it has proven to perform well in earlier studies \cite{fossez16_1793,fossez17_1927,jaganathen17_1974,jones17_1973}; our objective is to show how a simple, well established Hamiltonian based on effective scale arguments can capture the complex energy relations within the neutron-rich helium chain.

The one-body model space is the $spd$ space, built on the s.p. poles ${ 0{p}_{3/2} }$ and ${ 0{p}_{1/2} }$ and associated continua, each made of three segments in the complex momentum plane defined by the points ${ (0.2,-0.1) }$, ${ (0.4,0.0) }$ and ${ (6.0,0.0) }$ (all in ${ \text{fm}^{-1} }$) for the ${ {p}_{3/2} }$ partial wave, and the points ${ (0.25,-0.2) }$, ${ (0.5,0.0) }$ and ${ (6.0,0.0) }$ (in ${ \text{fm}^{-1} }$) for the ${ {p}_{1/2} }$ partial wave. The continuum associated with the ${ {s}_{1/2} }$ partial wave is real and defined by the points 0.1, 0.2, and 6.0\,${ \text{fm}^{-1} }$. Each segment defining the $s$ and $p$ continua are discretized with 12 points. Additionally, the ${ 1{s}_{1/2} }$ state was added to the s.p. basis for the $^{8-10}$He calculations by increasing the depth of the basis-generating WS potential, as its absence would make the identification of many-body states difficult. In fact, not including the ${ 1{s}_{1/2} }$ shell explicitly is possible and leads to identical results, but requires an unnecessary dense discretization of the ${ {s}_{1/2} }$ continuum to meet the unitarity condition. Finally, the ${ {d}_{3/2} }$ and ${ {d}_{5/2} }$ continua are represented by six harmonic oscillator shells each. We checked that adding higher partial waves only leads to an overall energy renormalization; hence, it does not change our results. It is worth noting that contrary to previous similar approaches \cite{volya05_470,michel04_93,jaganathen17_1974}, no truncations on the number of particles in the continuum are imposed in our work.
% and we consider a higher kmax cutoff...

%%%%%%%%%%%%%%%%%%%%%%%%%%%%%%%%%%% results %%%%%%%%%%%%%%%%%%%%%%%%%%%%%%%%%%%

\textit{Results}--Once the parameters of the WS potential have been optimized, there remains only one free parameter left, namely the strength of the spin-singlet central interaction ${V}_{c}$. We adjust ${V}_{c}$ for each energy of the known states in $^{6-8}$He and define the optimal value ${ {V}_{c}^{\text{(opt)}} }$ as the average over these values. In this way, we obtain ${V}_{c}^{\text{(opt)}}$=$-$5.709\,MeV with a standard deviation of $\sigma$=0.008\,MeV. The small value of $\sigma$ illustrates the ability of our model to describe the spectra of $^{6-8}$He. In fact, if we reduce the two-body interaction to the $^1S_0$ channel only and readjust the ground state of $^6$He to the experimental value, our predictions degrade only slightly. For instance, by considering the $0^+$ and $2^+$ states of $^{6}$He as well as the ${3/2}^-$ and ${5/2}^-$ states of $^{7}$He, the rms error on the energy is about 9 keV with the original interaction ($L \geq 0$), while it is about 26 keV with the simplified interaction ($L=0$). This demonstrates the dominant role of the $^1S_0$ interaction channel as expected from halo-EFT. By defining ${ {V}_{c}^{\text{(opt)}} }$ within a range constrained by known data, we ensure that if our model reproduces experimental data well, the parameter range is small and predictions are precise. This is analogous to the halo-EFT approach where effects of neglected higher-order terms are absorbed in the coupling constants of the model and the associated error. If the explicit three-body and higher-body forces were crucial, this would affect our ability to precisely reproduce experimental data.
% for 6He(0+,2+) and 7He(3/2-,5/2-):
%
% rms(E) = 0.0091 MeV [original]
% rms(E) = 0.0257 MeV [L=0, readjusted on 6He g.s.]

We wish to point out that the ${ {J}^{\pi} = {2}_1^{+} }$ state of $^{6}$He requires an abnormally large interaction strength (${V}_{c} \approx -6.8$\,MeV) to reproduce the experimental value; hence, is not included in the calculation of ${V}_{c}^{\text{(opt)}}$. The reason for this discrepancy (around 180\,keV) is the dominant $0{p}_{3/2} \rightarrow 0{p}_{1/2}$ structure of this state \cite{papadimitriou11_277}. In fact, the deviation between the calculated and experimental values for the ${2}_1^{+}$ state can be significantly reduced by slightly changing the strength of the spin-orbit term of the core-neutron potential. In this work, however, we decided to keep the one-body Hamiltonian fixed throughout.

In general, decay widths are not computed as accurately as energies. Moreover, energies and widths are highly correlated. For these reasons, we decided not to include decay widths when computing the energy uncertainty associated with ${V}_{c}$: $\Delta E = 0.5\left| E ({V}_{c}^{\text{(opt)}} + \sigma ) - E ({V}_{c}^{\text{(opt)}} - \sigma ) \right|$. We only consider a $1\sigma$ deviation because this is the minimal requirement to reproduce all known energies. The G-DMRG results for the energy spectra of the neutron-rich helium chain using ${ {V}_{c}^{\text{(opt)}} }$ are shown in Fig.~\ref{spectrum} and listed in Table~\ref{tab1}. In principle, there are also uncertainties coming from the core potential, but they were shown to be negligible as compared to the uncertainties coming from the valence-space interaction in Ref.~\cite{jaganathen17_1974}. Only a complete uncertainty quantification study (e.g., through a Bayesian analysis) could provide full theoretical uncertainties.

\begin{table}[htb]
	\caption{Experimental \cite{ensdf} and calculated energies with respect to the $^4$He g.s. (in MeV) and widths (in keV) for $^{5-10}$He. The uncertainty $\Delta E$ on energies (in MeV) is given in the last column.}
	\begin{ruledtabular}
		\begin{tabular}{cclllll}
			Nucleus		& ${ {J}^{\pi} }$	& ${ E_{\text{exp}} }$	& ${ {\Gamma}_{\text{exp}} }$	& ${ E_{\rm th} }$	& $\Gamma_{\rm th}$	& ${ \Delta E }$ \\
			\hline \\[-6pt]
			$^{5}$He 		& ${ {3/2}^{-} }$	& 0.798			& 648				& 0.766		& 671		& \\
			& ${ {1/2}^{-} }$	&			&				& 2.197		& 5903		& \\
			\hline \\[-6pt]
			$^{6}$He		& ${ {0}^{+} }$		& $-$0.972		&				& $-$0.974	&		& 0.006 \\
			& ${ {2}^{+} }$		& 0.824			& 113				& 1.007		& 207		& \\
			\hline \\[-6pt]
			$^{7}$He		& ${ {3/2}^{-} }$	& $-$0.527		& 150				& $-$0.507	& 142		& 0.007 \\
			& ${ {1/2}^{-} }$	&			&				& 0.844		& 2150		& 0.006 \\
			& ${ {5/2}^{-} }$	& 2.393			& 1990				& 2.344		& 1726		& 0.002 \\
			\hline \\[-6pt]
			$^{8}$He		& ${ {0}^{+} }$		& $-$3.10			&				& $-$3.176	&		& 0.014 \\
			& ${ {2}^{+} }$		& 0.0			& 600				& 0.116		& 776		& 0.009 \\
			\hline \\[-6pt]
			$^{9}$He		& ${ {1/2}^{+} }$	&			&				& $-$3.05		& 76		& 0.015 \\
			& ${ {1/2}^{-} }$	&			&				& $-$2.71		& 210		& 0.017 \\
			& ${ {5/2}^{+} }$	&			&				& 0.55		& 250		& \\
			\hline \\[-6pt]
			$^{10}$He 	& ${ {0}^{+} }$		&			& 				& $-$3.21		& $ < $ 1 keV	& 0.014
		\end{tabular}
	\end{ruledtabular}
	\label{tab1}
\end{table}

Consistently with Refs.~\cite{michel04_93,volya05_470,baroni13_753,baroni13_385}, we predict very broad ${1/2}_1^{-}$ states in $^{5,7}$He; these resonant states cannot be considered as genuine nuclear states because of their short lifetimes, see Refs.~\cite{fossez17_1916,fossez16_1335} for more detailled discussions. For $^{8}$He, we found that its g.s. has a complex structure~\cite{keeley07_1926,skaza07_1925}, with ${ {p}_{3/2} }$ and ${ {p}_{1/2} }$ occupations being about 2.58 and 0.18, respectively, and the remaining occupations (0.24) shared between the $s$ and $d$ partial waves. For comparison, the first excited ${2}^{+}$ state of $^{8}$He has ${ {p}_{3/2} }$ and ${ {p}_{1/2} }$ occupations of almost 3.0 and 1.0, respectively. This is reminiscent of the situation in $^{6}$He, whose g.s. has a strong dineutron component and the excited state has predominantly a particle-hole structure \cite{papadimitriou11_277}. 

%We note that the present model cannot provide precise charge and neutron radii without corrections accounting for effects beyond a static-core plus valence-neutron picture. Indeed, as shown in Ref.~\cite{papadimitriou11_277}, the ``core-swelling" effect (core polarization due to the valence neutrons) plays an important role in the charge radius budget.
We note that the present model cannot provide precise charge and neutron radii without corrections accounting for effects beyond a static-core plus valence-neutron picture, as for instance the ``core-swelling" effect (core polarization due to the valence neutrons) \cite{papadimitriou11_277}.

For $^{9}$He, we predict a narrow ${ {J}^{\pi} = {1/2}^{+} }$ g.s. and a close-lying ${ {J}^{\pi} = {1/2}^{-} }$ resonance with a larger width (these states could not be distinguished in the recent Gamow shell model study \cite{jaganathen17_1974} within statistical uncertainties), as well as a ${ {J}^{\pi} = {5/2}^{+} }$ resonance at higher energy. The uncertainty on the ${ {J}^{\pi} = {5/2}^{+} }$ state could not be estimated due to the instability of calculations for extreme values of ${V}_{c}$. These results are in relative agreement with experimental data from $(d,p)$ reactions \cite{kalanee13_1909}, and at variance with the study of isobaric analog states in $^{9}$Li \cite{uberseder16_1920}, as well as the no-core shell model with continuum calculations of Ref.~\cite{vorabbi18_1977} where the g.s. is predicted to have ${ {J}^{\pi} = {1/2}^{-} }$ and the first excited state to be a broader ${ {J}^{\pi} = {3/2}^{-} }$ resonance. We note that in Ref.~\cite{vorabbi18_1977}, in which only the two-body part of the normal-ordered three-body forces was considered, the $2^+_1$ state of $^{8}$He used to build the ${ {}^{8}\text{He} + n }$ channels was calculated as a bound state, and the only decay channel considered (for a fairly small number of channels) was one-neutron emission. Earlier quantum Monte Carlo results \cite{nollett12_835} stated the possible existence of a virtual ${ {J}^{\pi} = {1/2}^{+} }$ state in $^{9}$He, seen as a $\ell=0$ single-particle state above $^{8}$He, and a possible ${ {J}^{\pi} = {1/2}^{-} }$ state at higher energy (3-4 MeV). In our calculations, the ${ {J}^{\pi} = {1/2}^{+} }$ state in $^{9}$He is predicted to be a many-body resonance built almost entirely of excitations to the ${ {s}_{1/2} }$ continuum, but not a virtual state, see below.

We also make a prediction for the g.s. of $^{10}$He, which is calculated at an energy that is slightly lower than the g.s. of $^{8}$He. Taking into account the uncertainty on the g.s. energies of $^{8-10}$He, and the decay width of the g.s. of $^{9}$He, both one- and two-neutron decay channels are theoretically possible. Interestingly, it appears that the ground states of $^8$He, $^9$He, and $^{10}$He have almost identical partial-wave decompositions except for the ${ {s}_{1/2} }$ occupations, which are almost exactly zero, one, and two, respectively. In comparison, the $ {1/2}^{-}$ state of $^{9}$He is almost entirely built of the ${ {p}_{1/2} }$ component. The interplay between ${ {s}_{1/2} }$ and ${ {p}_{1/2} }$ continuum states is believed to be a determining factor for the phenomenon of parity inversion in $^{9}$He \cite{hansen01_1976}.

The present results shed new light on the nature of $^{9,10}$He. Early on, it was proposed using a three-body model \cite{aoyama02_1923} that the ground state of $^{10}$He might be a low-energy resonance ($E=0.05$\,MeV, $\Gamma=0.21$\,MeV) dominated by $s$ waves, suggesting that the reported observations of higher-energy  resonances at $\sim$1.8\,MeV \cite{kohley12_1215}) might in fact correspond to the first excited state of $^{10}$He. Concurrently, other studies \cite{barker04_2000,grigorenko08_1921} looked at the consistency between possible narrow ground state in $^9$He and a broad ground state at 1-3 MeV in $^{10}$He, and concluded that either the ground state of $^{10}$He has not been observed yet, or the $s$-wave scattering length in $^9$He must be less attractive. It was also suggested that the observed state in $^{10}$He might, in fact, corresponds to several states \cite{sharov14_1911,fortune15_1869}.
While the present study does not address the excited states of $^{10}$He, it goes beyond the limited three-body picture and supports the idea of a narrow ground state of $^{10}$He dominated by $s$ waves, built on the ${1/2}^+$ ground-state resonance of $^{9}$He. Due to the variability of approaches, different effective Hamiltonians, as well as the lack of uncertainty quantification, it is difficult to make a quantitative comparison with the previous theoretical studies.

The almost identical energies and partial-wave occupations of the ground states of $^{8}$He and $^{10}$He support the ${ {}^{8}\text{He} + 2n }$ cluster picture of $^{10}$He, in which an extended dineutron structure is present atop the four-neutron halo in $^{8}$He. In other words, $^{10}$He is predicted to be on the brink of forming a nuclear double-halo structure ($^{4}$He + $4n$ + $2n$) if not for a few tens of keV, similarly to the known ${ {}^{3}\text{He}{}^{4}\text{He}_{2} }$ trimer \cite{voigtsberger14_2016}. %In this context, we note similar energy relations for ${}^{8,9,10}$He and the dripline oxygen isotopes $^{26,27,28}$O \cite{fossez17_1927,jones17_1973}. Namely, $^{10}$He and $^{28}$O, both nominally doubly-magic, are predicted to have a threshold character and are expected to decay predominantly \textit{via} the $2n$ channel. The unbound nuclei $^{27}$O and $^9$He have complex spectra consisting of positive and negative parity states. In both neutron-rich oxygen and helium chains, the continuum space affects the energy relations in a profound way.

%%%%%%%%%%%%%%%%%%%%%%%%%%%%%%%%%%% conclusions %%%%%%%%%%%%%%%%%%%%%%%%%%%%%%%%

\textit{Conclusions}--In this study, we demonstrated that the intricate energetic relations within the neutron-rich helium chain ($^{5-10}$He) can be precisely described in a very large continuum space, by using a simple Hamiltonian justified by effective scale arguments. In the present $^4$He-plus-$(A-4)$\,neutron G-DMRG framework, the Hamiltonian reduces to a core-neutron potential optimized to the low-energy ${ n - ^{4}\text{He} }$ scattering, and a single central valence-neutron interaction term in the spin-singlet channel. The success of our approach can be understood in terms of the halo effective field theory approach to dilute systems.

Our calculations have, for the first time, no truncation on the number of particles in the continuum within the $spd$ model space. This milestone was enabled by using the G-DMRG method for open quantum systems in a basis of natural orbitals. In this way, we were able to consider the largest ever continuum space when making predictions for extremely neutron-rich threshold systems $^{9,10}$He.

By optimizing the single active parameter of our model, the strength of the two-body isoscalar central potential, we were able to reproduce known energy levels in $^{6-8}$He within tens of keV. We predict a parity inversion in $^{9}$He, which is a robust feature of our model, and showed that the ground states of $^{8-10}$He have almost identical $\ell$-content except for the $s$-wave. We predict $^{10}$He to be a threshold system, most likely a two-neutron emitter, but considering current theoretical uncertainties we cannot exclude the sequential two-neutron and direct one-neutron decay branches. The next generation of experimental studies will hopefully determine whether or not the ground state of $^{10}$He shows elements of a double-halo structure.

In conclusion, this work offers a way to revisit phenomenological approaches through a parameter reduction guided by effective scale arguments, providing a practical and reliable alternative to more complex approaches such as halo effective field theory or full-fledged $A$-body calculations for drip-line nuclei. The strategy outlined in this work could be easily exported to other approaches and physical systems. Additionally, it was brought to our attention that the Hamiltonian developed in this work can be seen approximately as the leading-order Hamiltonian of halo effective field theory (EFT) plus a perturbation. This idea was already formulated in the context of the nuclear interaction in Ref.~\cite{konig17_1986}, which shows that nuclei can be described using the leading order of an EFT approach in the unitary limit plus a small perturbation. In both cases, this strategy leads to great simplifications while still providing surprisingly precise results. The present study suggests that while the current halo-EFT strategy quickly increases the complexity of the Hamiltonian when using power counting rules, there might be alternative ways to develop simple and consistent effective descriptions of neutron-rich systems.

\begin{acknowledgments}
	 We thank Nicolas Michel for sharing the codes used to generate interaction matrix elements and optimize the core potential. We also thank Heiko Hergert, Scott Bogner and S. K\"onig for many useful comments, as well as Yannen Jaganathen and Erik Olsen for discussions. This work was supported by the U.S.\ Department of Energy, Office of Science, Office of Nuclear Physics under award numbers DE-SC0017887, DE-SC0013365 (Michigan State University) and DE-SC0018083 (NUCLEI SciDAC-4 collaboration), and by the National Science Foundation under award number PHY-1403906. An award of computer time was provided by the Institute for Cyber-Enabled Research at Michigan State University, and part of the computations was performed on local resources at Chalmers University of Technology supported by the Swedish Foundation for International Cooperation in Research and Higher Education (STINT, IG2012-5158).
\end{acknowledgments}

%\bibliographystyle{apsrev4-1}
%\bibliography{apsrev_refer_new}

%merlin.mbs apsrev4-1.bst 2010-07-25 4.21a (PWD, AO, DPC) hacked
%Control: key (0)
%Control: author (72) initials jnrlst
%Control: editor formatted (1) identically to author
%Control: production of article title (-1) disabled
%Control: page (0) single
%Control: year (1) truncated
%Control: production of eprint (0) enabled
%

\end{document}